\documentclass[preprint,nofootinbib]{revtex4}%
\usepackage{amssymb}
\usepackage{amsfonts}
\usepackage{amsmath}
\usepackage{amsmath}
\usepackage{graphicx}
\usepackage[usenames]{color}%
\usepackage{epsfig}
\providecommand{\U}[1]{\protect\rule{.1in}{.1in}}

\providecommand{\U}[1]{\protect\rule{.1in}{.1in}}
\definecolor{blue}{rgb}{0,0,1}

\definecolor{red}{rgb}{1,0,0}

\begin{document}
\title{Quark-Antiquark Potential as a Probe for Holographic Phase Transitions}
\author{Andr\'{e}s Anabal\'{o}n$^1$, Mariano Chernicoff$^2$, Gaston Giribet$^3$, Julio Oliva$^1$, Martín Reyes$^2$}

\affiliation{$^1$Departamento de F\'{\i}sica, Universidad de Concepci\'{o}n, Casilla, 160-C,
Concepci\'{o}n, Chile.}

\affiliation{$^2$Departamento de F\'{\i}sica, Facultad de Ciencias, Universidad Nacional Aut\'{o}noma de M\'{e}xico,   A.P. 70-542, CDMX 04510, M\'{e}xico.}

\affiliation{$^3$Center for Cosmology and Particle Physics, Department of Physics, New York University,  726 Broadway, New York, NY10003, USA.}

\begin{abstract}
In the recent paper \cite{Anabalon}, a higher-order phase transition between the planar, charged, 5-dimensional Reissner-Nordstr\"om-Anti-de Sitter black hole and a hairy black hole solution of the type IIB supergravity was investigated. Here, following a bottom-up approach, we set out to investigate these two phases of the theory by means of the holographic probe that describes a quark-antiquark in the dual gauge theory. We ask ourselves whether studying the quark-antiquark potential suffices to detect the change of behavior at different values of the parameter that controls the phase transition, this parameter being the ratio between the chemical potential and the temperature. We show that, while evaluating the probe on both phases leads to the same value at the point where the transition takes place, there is always one phase that dominates over the other with regard to this observable. The same can be said about higher-dimensional probes such as those involved in the computation of holographic entanglement entropy.
\end{abstract}
\maketitle

\section{Introduction}
The holographic correspondence \cite{Malda} provides a robust framework for exploring the dynamics of strongly coupled systems, particularly gauge theories in various settings. One of the most intriguing applications of this duality is its ability to shed light on the properties of the so called quark-gluon plasma, a state of matter observed in heavy-ion collisions where quarks and gluons interact strongly, behaving more like a fluid than a weakly coupled gas. In this context, holography has been instrumental in understanding the non-perturbative dynamics of such systems by mapping them to the corresponding gravitational duals.

In this work, we explore the phase structure of a system dual to a 5-dimensional asymptotically Anti-de Sitter (AdS) charged black hole, which serves as the holographic dual of a strongly coupled quark-gluon plasma with a finite chemical potential in $\mathcal{N}=4$
super Yang-Mills theory for $SU(N)$ at large $N$ and large 't Hooft coupling $\lambda$. Our analysis focuses on a recently identified third-order phase transition between two distinct plasma phases, which is manifested as a transition between a standard Reissner-Nordstr\"om-AdS$_5$ black hole and a hairy black hole solution \cite{Anabalon} that can be regarded as a limit of solutions studied in \cite{Cvetic}. This transition occurs at a critical ratio of chemical potential to temperature $\psi\equiv \mu/(2\pi T)=1$, marking the emergence of a new phase characterized by a scalar hair that modifies the thermodynamics of the system. For a ten-dimensional perspective on these ideas, see \cite{Henriksson:2019zph}.

Here, we are going to investigate these two phases using a particular string configuration which is dual to the quark-antiquark potential in the gauge theory \cite{Malda2}. To be more precise, we will show that by computing the $q \overline{q}$-potential using the expectation value of a rectangular Wilson loop, we can gain insight into how the phases differ in their ability to confine or screen charges. In the Reissner-Nordstr\"om-AdS black hole phase, the quark-antiquark potential exhibits a Coulomb type behavior at short separations, consistent with a deconfined plasma. As the system undergoes a phase transition to the hairy black hole phase, the potential partially reflects the new phase structure: while evaluating the probe on both phases leads to the same value at the point where the transition takes place, namely $\psi =1$, there is always one phase that dominates over the other with regard to this observable. 

\section{The quark-antiquark potential}

In the context of the holographic correspondence \cite{Malda}, the computation of the static quark-antiquark potential provides a direct way of probing the dynamics of strongly coupled gauge theories. The expectation value of a Wilson loop in the gauge theory is given by the on-shell action of a fundamental string in the corresponding gravitational background \cite{Malda2, Yonenyay}. Namely, the quark-antiquark potential is obtained from the expectation value of a Wilson loop
\begin{equation}
\langle W(\mathcal{C})\rangle\thicksim e^{{-\text{T} E(L)}},
\end{equation}
where $\mathcal{C}$ is a rectangular loop with a spatial extent $L$ and a lapse $\text{T}$. This Wilson loop is dual to the normalized on-shell action of a fundamental U-shape string whose endpoints are fixed on the AdS$_5$ boundary and separated by a distance $L$. The string extends into the bulk, probing the spacetime geometry, and in the context of  the holographic correspondence, this provides information about the strongly coupled gauge theory. 

To study the string dynamics, we need to consider the Nambu-Goto action\footnote{We have set the tension of the string to $1/(2\pi)$.}
\begin{equation} \label{ng}
S= -\frac{1}{2\pi}\int{d\tau
d\sigma}\,\sqrt{-\det{g_{\alpha\beta}}} \,,
\end{equation}
where $g_{\alpha\beta}\equiv
G_{\mu\nu}(X)\partial_{\alpha}X^{\mu}\partial_{\beta}X^{\nu}$ is the induced worldsheet metric ($\mu , \nu =0,1,2,3,4$), where $\sigma^{\alpha}\equiv (\tau, \sigma)$ ($\alpha , \beta = 0,1$) and where $G_{\mu\nu}(X)$ is the spacetime metric. Although the computation of the probe in the context of string theory demands the inclusion of the coupling of the string with the gauge fields that are generated by dimensionally reducing from 10 to 5 spacetime dimensions; however, here we will consider the minimal setup and analyze the standard Nambu-Goto action (\ref{ng}) for the string probe in 5 dimensions. In fact, we have performed the computation including the coupling with the gauge fields and the qualitative behavior is the same. For simplicity, we focus on action (2). 

Choosing a generic static gauge where $\tau=t$ and $\sigma=x$, the profile of the string embedding can be described by $X^{\mu}=(t,r(x),x,0,0)$, with appropriate boundary conditions. The on-shell action computed in this way is actually divergent. This is due to the infinite energy density of the U-shape string. To obtain a physically meaningful result a regularization is required. This is typically achieved by subtracting the on-shell action evaluated on the configuration corresponding to two straight strings extending from the boundary to the horizon. The regularized energy obtained in this way represents the binding energy of the quark-antiquark pair in the thermal plasma. From the dual gauge theory point of view, this regularization can be thought of as the subtraction of the self-energy of the isolated quarks. The resulting expression for $E(L)$ thus represents the potential energy associated with the quark-antiquark pair in the presence of a thermal medium as a function of the separation $L$. Generically, at short distances $L\ll 1/T$, with $T$ being the temperature, the quark-antiquark potential behaves similarly to the zero-temperature case, showing a Coulomb like behaviour,
\begin{equation}
E(L)\sim -\frac{C}{L} \ ,
\end{equation}
with $C$ being a constant related to the 't Hooft coupling, $\lambda$. As the separation $L$ increases, the U-shape string, dual to the $q\overline{q}$-potential, dips further into the bulk where thermal effects become significant and the potential shows screening behavior due to the thermal plasma. For distances $L\gg 1/T$, the interaction between the quark and antiquark weakens, and beyond a critical separation $L_c$ the bound state becomes unstable, indicating the quarks are no longer confined.

\section{Probing the plasma-plasma phase transition}

The computation reviewed above can be carried out for the scenario described in \cite{Anabalon}. It involves a hairy black hole, which serves as the dual description of a strongly coupled fluid resembling a quark-gluon plasma. From the bulk perspective, the hairy phase is a solution to the STU model, akin to the Reissner-Nordstr\"om-AdS$_5$ black hole. Notably, a phase transition occurs between these two geometries. This happens at the point\footnote{Some authors use the normalization $\mu = 2\pi \sqrt{2}T$.} $\mu = 2\pi T$, with $\mu$ being the chemical potential and $T$ the temperature.

The new phase (hereafter referred to as hairy phase) is described by the metric
\begin{equation}\label{Hairymetric}
    ds^2 = \Omega(x) \left( -F(x) L^2 dt^2+ \frac{dx^2}{4x^2 (x-1)F(x)\eta} + dy^2 + dz^2 +dw^2  \right),
\end{equation}
where the metric functions are given by
\begin{equation}
    \Omega(x)=\frac{x^{2/3}\eta}{x-1}, \hspace{2cm} F(x) = L^{-2}-\frac{(-1+x)^2(q_1^2 - q_2^2 x)}{\eta x^2};
\end{equation}
with
\begin{equation}
    q_1 = \frac{\mu x_0}{x_0 -1}, \hspace{1cm} q_2= -\frac{\mu}{x_0-1}\ ,
\end{equation}
and
\begin{equation}
    \eta = \frac{\mu^2 L^2 (x_0-1)}{x_0}\ ,
\end{equation}
where $x_0$ is determined by $F(x_0)=0$.

It is convenient to define the dimensionless parameter $\psi >0$ in terms of the boundary quantities $T$ and $\mu$, such that
\begin{equation}
    x_0 = \left(\frac{\mu}{2\pi T} \right)^2 \equiv \psi^2.
\end{equation}
This means that the phase transition occurs at $\psi=1$.
Also note that, in this particular set of coordinates, the spacetime geometry described by (\ref{Hairymetric}), has the conformal boundary located at $x=1$. In what follows, and in order to simplify our numerical computation, we will work with a set of coordinates in which the boundary is located at infinity, this is achieved through the coordinate transformation given by
\begin{equation}
    x = 1+ \frac{4\pi(\psi^2 -1)}{r^2} + \frac{32 \pi^2 (\psi^2-1)^2}{3r^4} + \frac{64\pi^6(\psi^2-1)^3}{3r^6} + \mathcal{O}(r^{-8}),
\end{equation}
with $r$ being dimensionless. A similar procedure must be applied to the planar Reissner-Nordstr\"om-AdS$_5$ metric in order to compare the results; namely,
\begin{equation}\label{RNmetric}
    ds_{RN}^2 = r^2T^2 \left(-H(r)dt^2+ \frac{dr^2}{H(r)} + dy^2+dz^2 + dw^2  \right),
\end{equation}
with the function 
\begin{equation}
H(r)=1-\frac{\pi^4 (1+\sqrt{1+8\psi^2})^2(1+12\psi^2+\sqrt{1+8\psi^2})}{8r^4}+\frac{\pi^6\psi^2 (1+\sqrt{1+8\psi^2})^4}{4r^6}.
\end{equation}

As mentioned above, calculating the quark-antiquark potential reduces to determining the area of the worldsheet configuration whose boundary corresponds to the temporal Wilson loop. Parameterizing the static string worldsheet as follows
\begin{equation}
     X^{\mu} = (t(\tau),r(\sigma),y(\sigma),0,0),
\end{equation}
the Nambu-Goto action for the metric \eqref{Hairymetric} becomes
\begin{equation}\label{NG-hairy}
    S =\frac{1}{2\pi } \int d\tau d\sigma \ \Omega(r) \dot{t}(\tau ) \sqrt{
    L^2 F(r) y'(\sigma )^2+  \frac{3 r^4 \,P^2(r^2,\psi^2)\, r'(\sigma )^2}
 {T^2 Q(r^2,\psi^2)R^2(r^2,\psi^2)}
  },
\end{equation}
with the polynomials
\begin{eqnarray}
P(x,y)&=& 3x^2+16\pi^2 \,x(y-1)+48\pi^4\, (y-1)^2\ ,\\
Q(x,y)&=& 3x^2+8\pi^2 \,x(y-1)+16\pi^4\, (y-1)^2\ ,\\
R(x,y)&=& 3x^3+12\pi^2\, x^2(y-1)+ 32\pi^4\, x(y-1)^2+64\pi^6\, (y-1)^3 \ .
\end{eqnarray}

The absence of explicit $y$-dependence yields a conserved quantity $\Pi_y =\frac{\partial \mathcal{L}}{ \partial y'(\sigma)} = C_1$ being a constant. Using this, the explicit dependence of $y'$ in terms of $C_1$ can be determined; namely,
\begin{equation}\label{yaux}
    y'(\sigma) =  
    \frac{ \sqrt{3}\, r'(\sigma)\, C_1r^2P(r^2,\psi^2 )}
        {T\, R(r^2,\psi^2 )\, \sqrt{ F(r) \left(F(r)\Omega^2(r) \dot{t}^2-C_1^2\right) Q(r^2,\psi^2 ) }}\ .
\end{equation}
From this expression, one can identify the location of the turning point $r_0$ of the U-shape string, defined by $y'(\sigma)\rightarrow \infty$, and relate it to the constant $C_1$ through
\begin{equation}\label{c1}
    C_1 = \sqrt{F(r_0)} \Omega(r_0) \dot{t}(\tau)\ .
\end{equation}

Substituting \eqref{c1} into \eqref{yaux}, we obtain the dependence of $y'$ on $r_0$, whose integration yields the quark separation distance given by
\begin{equation}
    L = \frac{1}{T} 
    \int_{r_0}^{\infty} dr\, r^2 \, \frac{P(r^2,\psi^2 )\Omega(r_0)}{R(r^2,\psi^2 )}
\sqrt{\frac{12\,    
    F(r_0) }
    {F(r) \left(F(r) \Omega^2(r)-F(r_0) \Omega^2(r_0)\right)\, Q(r^2,\psi^2 )} 
    }\ .
\end{equation}

This procedure can be repeated for the asymptotically, locally AdS$_5$ Reissner-Nordstr\"om geometry \eqref{RNmetric} to determine the quark-antiquark separation for its dual plasma. 

Expression \eqref{yaux} can be plugged into the Nambu-Goto action \eqref{NG-hairy}, allowing to calculate the quark-antiquark potential as a function of the turning point $r_0$. The expression for the hairy phase takes the form
\begin{equation}\label{Potential-hairy}
    V_{q\bar{q}} =  \frac{1}{T}\int_{r_0}^{\infty} dr \, r^2\,
\frac{P(r^2,\psi^2)\Omega^2(r)}{R(r^2,\psi^2 )}
\sqrt{
\frac{12\, F(r)}{(F(r) \Omega^2 (r)-F(r_0) \Omega^2(r_0))\, Q(r^2,\psi^2)}
} \ .
\end{equation}

As explained before, the integral (\ref{Potential-hairy}) is infinite and must be regularized. This is done in the standard way by subtracting the mass of two free static quarks \cite{Yonenyay}. The calculation can be reproduced for the metric \eqref{RNmetric}. 

\begin{figure}
    \centering    \includegraphics[width=0.6\linewidth]{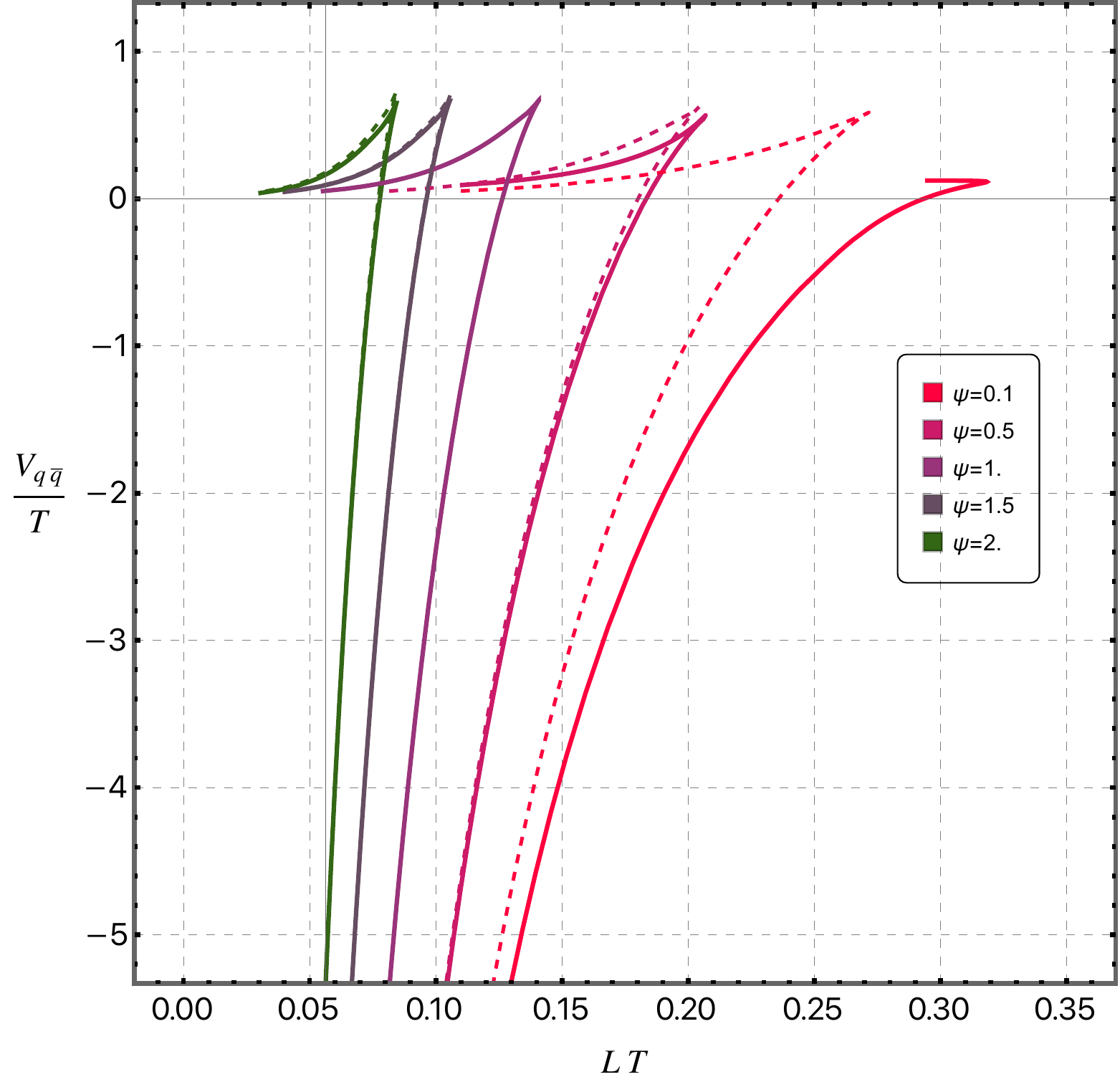}
    \caption{The quark-antiquark potential $V_{q\bar{q}}$ as function of the pair separation $L$ for different values of $\psi$. The numerical results for the hairy and Reissner-Nordstr\"om scenarios are shown in solid and dashed lines, respectively.}
    \label{fig:V vs L}
\end{figure}

In order to study the behavior of the quark-antiquark potential $V_{q\bar{q}}/T$ as a function of the separation $LT$, we have relied on computational methods to obtain our results. First, the parametric analysis, in terms of the implicit parameter $r_0$, is shown in Fig. \ref{fig:V vs L}, with solid lines for the hairy phase and dashed lines for the Reissner-Nordstr\"om-AdS$_5$ black hole. At large $\psi $, where the new phase is seen to emerge, there are string configurations that are favored, in the sense of having smaller free energy. In fact, we see the special behavior that the string configuration exhibits around $\psi =1$, where the evaluation of the probe for both the hairy and the Reissner-Nordstr\"om solutions coincides. However, we observe that the string probe does not directly detect the region of the parameter space where the phase transition takes place as one of the two phases always dominate for $\psi \neq 1$.   
\begin{figure}
    \centering
\includegraphics[width=0.7\linewidth]{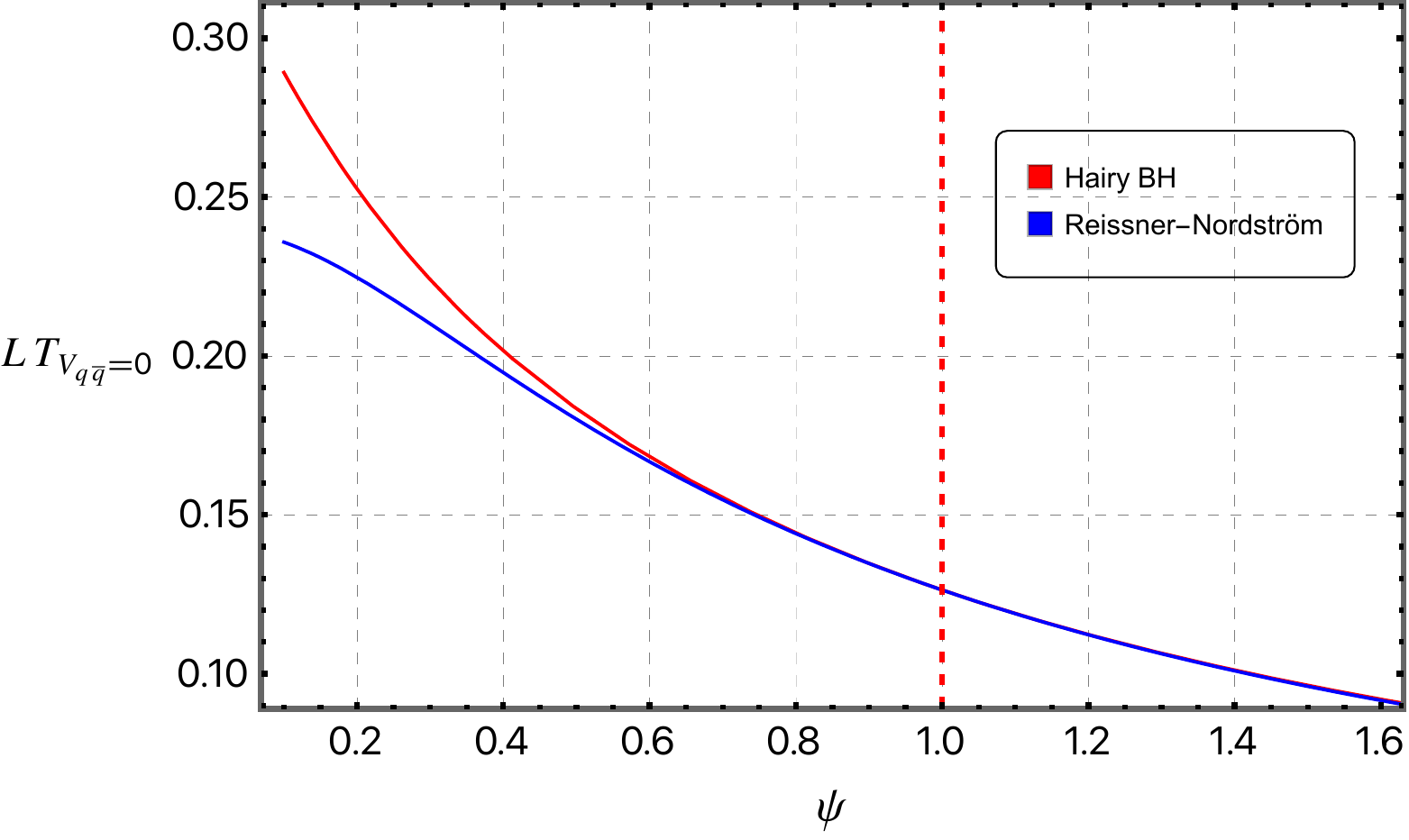}
    \caption{The separation of the pair $L$ at which the potential vanishes as function of the parameter $\psi$. In red the behavior of the hairy phase, while in blue the corresponding for the Reissner-Nordstr\"om is shown.}
    \label{fig:LTnull vs psi}
\end{figure}

The results presented in Fig. \ref{fig:V vs L} and Fig. \ref{fig:LTnull vs psi} show the following: the behavior of $V_{q\bar{q}}$ as a function of the quark-antiquark separation $L$ is consistent with a plasma-plasma phase transition when $\psi = 1$, in the sense that the observable evaluated on the hairy black hole and on the Reissner-Nordstr\"om black hole coincides when $\psi = 1$. Outside $\psi =1 $, in contrast, one of the branches always dominate over the other.

The case of higher-codimension probes using the holographic entanglement entropy \cite{Ryu} can also be studied. For example, one can consider the boundary-anchored slab-type configuration with translational symmetry in the $y$ direction and follow the prescription presented in \cite{Kundu} to regularize the corresponding holographic entanglement entropy. For this case, one finds results similar to those of the probe case we analyzed above.

\subsection*{Acknowledgements}

We thank Carlos Núñez and Ricardo Stuardo for enlightening comments. M.C., G.G., and J. O. thank the people at the campus in Santiago de Chile of Universidad de Concepci\'on for their hospitality during their visit.
The work of M.R. is supported by CONAHCyT Ph.D. fellowship. The work of M.C. and M.R. is partially supported by Mexico’s National Council of
Humanities, Sciences, and Technologies (CONAHCyT) grant A1-S-22886, and DGAPAUNAM grant IN116823. The work of J.O. and A.A. is partially supported by FONDECYT grants 1210635, 1221504,
1230853 and 1242043.

\end{document}